\documentclass[11pt]{article}   
\usepackage{graphicx,amssymb}    
\usepackage{epsfig}
\usepackage{graphics}     
\usepackage{psfrag}     
\usepackage{epsfig}\parskip 5pt plus 1pt
\usepackage{bbm}
\usepackage{amsmath}
\usepackage{amssymb}
\usepackage{amsfonts}
\usepackage{mathrsfs}
\usepackage{graphicx}
\usepackage{color}
\usepackage{graphicx,amssymb}    
\usepackage{epsfig}
\usepackage{graphics}     
\usepackage{psfrag}

\def\cta{\cos\theta_A}

\def\sta{\sin\theta_A}

\def\tanb{\tan\beta}
\def\cotb{\cot\beta}  
\def\ma{m_{A_1}} 

\def\to{\rightarrow} 
\def\gam{\gamma}
\def\tauptaum{\tau^+\tau^-} 

\def\3mups{M_{\Upsilon(3S)}}
\def\mupmum{\mu^+\mu^-}
\def\Br{\rm Br}

\begin{document}      
\vspace*{-0.5in}      
\renewcommand{\thefootnote}{\fnsymbol{footnote}}      
\begin{flushright} 
LPT Orsay/10-54,~~
SINP/TNP/2010/10
\end{flushright}      
\vskip 5pt

\begin{center}      
{\Large {\bf 
A possible connection between neutrino mass generation 
and the lightness of a NMSSM pseudoscalar 
}} \\
\vspace*{1cm}
{\sf Asmaa Abada $^{1,}$\footnote{E-mail 
 address: {\tt asmaa.abada@th.u-psud.fr}}},  
{\sf Gautam Bhattacharyya  $^{2,}$\footnote{E-mail 
 address: {\tt gautam.bhattacharyya@saha.ac.in}}}, 
{\sf Debottam Das $^{1,}$\footnote{E-mail 
 address: {\tt debottam.das@th.u-psud.fr}}},  
{\sf C\'{e}dric Weiland $^{1,}$\footnote{E-mail 
 address: {\tt cedric.weiland@th.u-psud.fr}}}

\vspace{10pt} \small $^{1)}${\it Laboratoire de Physique Th\'eorique,
    Universit\'e de Paris-sud 11    \\
    B\^atiment 210, 91405 Orsay Cedex, France} \\
  $^{2)}${\it Saha Institute of Nuclear Physics,
    1/AF Bidhan Nagar, Kolkata 700064, India} \\

\normalsize      
\end{center} 
\vskip 10pt
 
\begin{abstract}   
  One of the interesting properties of the NMSSM is that it can accommodate a
  light pseudoscalar of order 10 GeV. However, such scenarios are challenged
  by several experimental constraints, especially those related to the
  fermionic decays of the pseudoscalar. In this Letter, we extend the NMSSM
  field content by two gauge singlets, with lepton numbers $+1$ and $-1$. This
  serves the twin purpose of generating neutrino masses via the inverse
  seesaw mechanism and keeping the option of a very light pseudoscalar
  experimentally viable by opening dominant invisible decay channels of the
  pseudoscalar which help it evade the existing bounds.
  
\vskip 0.4cm \noindent   
\texttt{PACS Nos:~ 12.60.Jv, 14.80.Cp, 14.60.St} \\   
\texttt{Keywords:~ Neutrino mass, Inverse seesaw, NMSSM}   
\end{abstract}

\renewcommand{\thesection}{\Roman{section}}
\setcounter{footnote}{0}
\renewcommand{\thefootnote}{\arabic{footnote}}


We consider the Next to Minimal Supersymmetric Standard Model (NMSSM)
which provides a natural solution to the so-called $\mu$ problem
through the introduction of a new gauge singlet superfield $\hat S$ in
the superpotential. In the NMSSM, the $\mu$ parameter is linked to the
vacuum expectation value (VEV) of the scalar component of $\hat S$
whose size is of the order of the supersymmetry breaking scale
\cite{NMSSM}.  Since it permits a scale invariant superpotential, the
NMSSM is the simplest supersymmetric generalization of the Standard
Model (SM) in which the supersymmetry breaking scale is the only mass
scale in the Lagrangian.  Moreover, by providing an additional tree
level contribution to the quartic term of the scalar potential it can
ameliorate the `little hierarchy problem' of the MSSM, which is
related to the requirement of a large ($\gg M_Z$) soft supersymmetry
breaking mass that can push the mass of the lightest neutral Higgs
beyond the LEP-2 limit of 114 GeV \cite{pdg}. Furthermore, the NMSSM
can admit a very light CP-odd Higgs boson ($m_{A_1}\sim 1 - 10$ GeV)
\cite{manirev,ulrichrev}. The main objective of this Letter is to
improve the experimental viability of such a light pseudoscalar by
providing it with a dominant invisible decay mode in a minimal
extension of the NMSSM which contains a source of lepton number
violation that also yields an acceptable neutrino mass.

In the NMSSM, the lightest CP-odd physical scalar $A_1$ can be decomposed as
\begin{equation}
\label{A1-def}
A_1 \equiv \cta A_{\rm MSSM} + \sta A_S \, ,
\end{equation}
where $A_{\rm MSSM}$ is the MSSM part of the CP-odd scalar, which arises
solely from the NMSSM Higgs doublets, and $A_S$ is the part that arises from
the new singlet superfield $\hat S$.  It is the singlet admixture, i.e. the
$\sta$ projection, that allows the NMSSM pseudoscalar to be much lighter than
what it could have been in the MSSM. On the other hand, if $A_1$ is very light
then its detection crucially depends on its couplings to quarks and leptons,
which depend on $\cta$.  These couplings can be extracted from the following
part of the Lagrangian \cite{gunion_dermisek}:
\begin{equation}
  \label{cabbdef}
  {\cal L}_{{A_1}f\bar f} = X_{u(d)} \frac{g m_f}{2M_W}
  \bar f \gamma_5 f A_1 \, ,
\end{equation} 
where $g$ is the SU(2) gauge coupling, $X_d (X_u) = \cta\tanb \ (\cta\cotb)$
for down-type (up-type) fermions, $\tan\beta \equiv v_u/v_d$ with $v_u$ and
$v_d$ denoting the up- and down-type Higgs VEVs.  

A light pseudoscalar is phenomenologically interesting mainly for two reasons:

\noindent ($i$) The direct search limit of 114 GeV on the mass of the SM-like
Higgs (a slightly smaller limit of 93 GeV for the MSSM neutral Higgs) is
obtained from its non-observation at the highest energy run at LEP-2, where
the Higgs was expected to be produced from gauge interactions (full strength
$ZZh$ coupling) and decay dominantly ($\sim 75\%$) into $b\bar b$ final
states. In the NMSSM, there could be two important changes:
\begin{itemize}
\item The lightest CP-even Higgs ($h$) may have a large singlet component,
  which leads to a significant dilution of the $ZZh$ coupling. The Higgs
  production cross section will then be reduced, and hence the lower limit on
  $m_h$ will be relaxed.

\item $h$ may dominantly decay into a pair of $A_1$, with each $A_1$ decaying
  into $f\bar f$, where $f$ is $b$, $\tau$, $c$, or $\mu$, depending on the
  kinematic thresholds. Therefore, the existing LEP-2 Higgs search strategy
  from $2b$ final states would fail as one should look for $4f$ final states.
  In fact, a reanalysis of the LEP-2 data by the LEP Collaboration has already
  put constraints on $\frac{\sigma(e^+e^- \rightarrow Zh)}{\sigma_{\rm
      SM}(e^+e^- \rightarrow Zh)} \times {\rm Br} (h\rightarrow A_1 A_1)\times
  {\rm Br}(A_1\rightarrow f \bar f)^2$, where $f=b$ or $\tau$ depending on
  kinematics thresholds \cite{Schael:2006cr,aleph}.  Similarly, for $\ma <
  2m_\tau$, upper limits have been placed on $\sigma(p \bar p \rightarrow
  hX)\times {\rm Br} (h\rightarrow A_1 A_1) \times {\rm Br} (A_1\rightarrow
  \mu^+ \mu^-)^2$ by the D0 collaboration at Fermilab Tevatron \cite{D0}.  The
  upshot is that the lower limit on $m_h$ is slightly weakened, and a smaller
  value of $m_h$ (e.g. 105 GeV) is in fact preferred by electroweak precision
  tests \cite{dermisekall}.

\end{itemize}

\noindent ($ii$) If the lightest supersymmetric particle (LSP) happens
to be very light (a few GeV), then a light $A_1$ offers the
possibility of $s$-channel LSP pair-annihilation into an on-shell
$A_1$. This resonance channel has a special significance when one
attempts to account for the observed dark matter relic abundance. It
has recently been shown that a light LSP of mass $\sim 10$ GeV can
have interesting consequences in the context of the recent DAMA/CoGeNT
results \cite{nmssmdarkall}.

Let us now briefly discuss the existing bounds on the mass of the light
pseudoscalar. The constraints on $X_d$, defined in Eq.~(\ref{cabbdef}), for
$\ma$ approximately in the range of 1 to 10 GeV have been summarized in
\cite{ulrichrev,gunion_dermisek}.  Measurements of $\Delta M_{d,s}$, ${\rm
  Br}(\bar B \to X_s \gamma)$, ${\rm Br}(B^+ \to \tau^+ \nu_\tau)$, and
particularly, ${\rm Br}(\bar B_s \to \mu^+ \mu^-)$ severely constrain
$m_{A_1}$ \cite{Domingo:2007dx}.  The rates of these processes primarily
depend on the choice of $\tan\beta$ and the soft supersymmetry breaking
trilinear term $A_t$, and the constraints are in general weaker when these
parameters are small. Values of $m_{A_1}$ between 1 GeV and $m_b$ are
generally disfavored from $B$-meson data \cite{ulrichrev}. Constraints on
$\ma$ also arise from radiative $\Upsilon$ decays
\cite{cleoiii,babar1,babar2}, namely, $\Upsilon (nS) \to \gamma A_1$, with
$A_1 \to \mu^+\mu^- (\tau^+\tau^-)$ (further investigated and reviewed in
\cite{gunion_dermisek}). Severe constraints also arise as a consequence of
$\eta_b -A_1$ mixing \cite{Drees:1989du,Domingo:2009tb,Domingo:2010am}.  The
different $\ma$ windows which are sensitive to different processes are listed
in Table \ref{cons}.  The table also shows the ranges where the LEP (ALEPH
\cite{aleph} and OPAL \cite{Abbiendi:2001kp}) constraints are applicable. The
origin of all these constraints can be traced to the visible decay modes of
$A_1$.

\begin{table}
\begin{center}
\hspace*{-0.5cm}\begin{tabular}[ht]{|c|c|c|c|c|}
\hline
\hline
{\small Processes} &{\footnotesize$m_{A_1}<2m_{\tau}$} &\footnotesize
$[2m_{\tau}$,$ 9.2$ GeV$]$ &
\footnotesize $[9.2$ GeV,$M_{\Upsilon(1S)}]$
&\footnotesize $[M_{\Upsilon(1S)}$,$2m_B]$ \\
\hline
\hline
{ { \small $\Upsilon \to \gam A_1 \to \gam + (\mupmum, gg, s \bar s)$
}} 
& $\checkmark$
& $\times$ & $\times$ & $\times$  \\
\hline
\hline
{ { \small $\Upsilon\to \gam A_1 \to \gam \tauptaum$ } }& $\times$ & 
$\checkmark$ &$\times$ & $\times$   \\
\hline
\hline
{{ \small $A_1$--$\eta_b$ mixing}} &$\times$  &$\times$  &$\checkmark$ &
$\times$     \\
\hline
\hline
{{ \small $e^+ e^- \rightarrow
 Z + 4\tau$ (ALEPH)}} &$\times$ & $\checkmark$  & $\checkmark$
& $\checkmark$    \\
\hline
\hline
{{ \small $e^+ e^- \rightarrow b \bar b \tauptaum$ (OPAL)}}  &$\times$  
&$\times$  &$\checkmark$ &$\checkmark$   \\
\hline
\hline
\end {tabular}
\caption {\em \small Different processes constraining different $\ma$ windows.
  The ``$\checkmark$'' symbol in a given entry attests the existence of 
  important or meaningful constraints from a given process, while the 
  ``$\times$'' symbol implies otherwise.}
\label{cons}
\end{center}
\end{table} 


However, the situation may dramatically change if $A_1$ has dominant invisible
decay modes.  Its decay into a pair of stable neutralinos (if kinematically
possible) is one such example.  The BABAR Collaboration \cite{invisible} at
the PEP-II $B$-factory has, however, searched for radiative $\Upsilon$-decays
where a large missing mass is accompanied by a monochromatic photon, and from
its non-observation has set a (preliminary) 90\% C.L. upper limit on ${\rm
  Br}(\Upsilon(3S) \to \gamma A_1) \times {\rm Br}(A_1 \to~{\rm invisible})$
at $(0.7-31)\times10^{-6}$ for $\ma$ in the range of 3 to 7.8 GeV.

In this Letter we further explore the possibility of invisible decay channels
that would allow a light $A_1$ escape detection even outside the range of 3 to
7.8 GeV.  We show that if we extend the NMSSM by two additional gauge singlets
with non-vanishing lepton numbers, they would not only provide a substantial
invisible decay channel of $A_1$ but, as a bonus, would also generate small
neutrino masses through lepton number violating ($\Delta L =2$) interactions.
The visible decay branching ratios of $A_1$ would then be reduced. As a
result, the constraints on $X_d$ would be weakened.  A light $A_1$ can then be
comfortably accommodated.


In the framework of the NMSSM, neutrino masses and mixings can be
generated via different mechanisms, either under the assumption of
$R$-parity conservation or $R$-parity violation (RpV).  In the latter
case, light neutrino masses/mixing data can be successfully
accommodated via the inclusion of explicit trilinear and/or bilinear
RpV terms~\cite{rpvnmssm1}, or through spontaneous violation of lepton
number in the presence of right-handed neutrino superfields (in
addition to the NMSSM singlet $\hat S$)~\cite{rpcnmssm1}.
Alternatively, one can employ a standard seesaw mechanism in a
$R$-parity conserving setup by adding three gauge singlet neutrino
superfields $\hat N_i$ to the NMSSM particle content
\cite{rpcnmssm2}. In this case, the light neutrino masses originate
from the tiny Yukawa couplings and (dynamically generated) TeV scale
Majorana mass terms.


Here, we follow none of the above paths.  We rather implement the
``inverse seesaw" mechanism \cite{inverse} by adding two gauge singlet
superfields with opposite lepton numbers ($+1$ and $-1$). During this
implementation we assume that $R$-parity is conserved, i.e. we do not
admit any $\Delta L = 1$ term in the superpotential.  To appreciate
the advantages of the inverse seesaw over the standard one we look at
the difference, from the point of view of effective operators, between
the properties of the dimension-5 (d-5) Weinberg operator which is
lepton number {\em violating} and the dimension-6 (d-6) operator which
is lepton number {\em conserving}. In the d-5 case, the light neutrino
mass is given by the seesaw formula $m_\nu \sim m_D^2/M$, where $m_D =
f v$ is a Dirac mass, with $v$ as the electroweak VEV and $f$ as a
generic Yukawa coupling. The source of lepton number violation is the
Majorana mass $M$. If we demand the Yukawa coupling $f$ to be order
one, then for $m_\nu$ to be $\sim$ 1 eV, $M$ has to be close to the
gauge coupling unification scale.  On the other hand, the d-6 operator
goes as $1/M^2$, but since this operator conserves lepton number, its
coefficient is not related to that of the d-5 operator
\cite{asmaa}. Therefore, for the d-6 case, one needs a separate source
of lepton number violation to generate light neutrino mass. The
generic form of this mass is $m_\nu \sim (m_D^2/M^2) \mu$, where $\mu$
is a lepton number violating ($\Delta L=2$) mass parameter. In this
case, one can comfortably keep the fundamental scale $M$ close to TeV,
i.e. within the LHC reach, and yet choose $f$ to be order unity which
can trigger large lepton flavor violation. Here, the lightness of the
neutrino mass (eV, or even lighter) is due to the smallness of $\mu$.
Having this small dimensionful term in the Lagrangian is technically
natural in the sense of 't Hooft \cite{thooft}, as in the limit of
vanishing $\mu$ one recovers the lepton number symmetry. The structure
of light neutrino mass obtained via inverse seesaw conforms to the
principle of `low fundamental scale, large Yukawa coupling and light
neutrino mass' \cite{Dudas}.

In our model, the superpotential is given by 
\begin{eqnarray}
W&=&W_{\rm NMSSM}+W'\ , ~~{\rm where,}
\label{Supot}
\end{eqnarray}
\begin{eqnarray}
W_{\rm NMSSM}&=&f^d_{ij} \hat H_d  \hat Q_i  \hat D_j 
              +f^u_{ij}  \hat H_u  \hat Q_i \hat U_j 
              + f^e_{ij}  \hat H_d \hat L_i  \hat E_j +
              \lambda_H  \hat S  \hat H_d  \hat H_u 
	      +\frac{\kappa}{3}  \hat S^3 \ , \\	
W' &=& f^\nu_{ij} \hat H_u \hat L_i  \hat N_j
         + (\lambda_N)_i \hat S \hat  N_i {\hat X_i}   
          + \mu_{Xi}  \hat X_i \hat X_i \, . 
         \label{model-exp}
\end{eqnarray}
In the above expressions, $\hat H_d$ and $\hat H_u$ are the down- and
up-type Higgs superfields; $\hat Q_i$ and $\hat L_i$ denote the SU(2)
doublet quark and lepton superfields; $\hat U_i$ ($\hat D_i$) and
$\hat E_i$ are the SU(2) singlet up (down)-type quark superfields and
the charged lepton superfields, respectively.  We have denoted the
Yukawa couplings by $f$ with appropriate flavor ($u,d,e,\nu$) and
generation indices ($i,j = 1,2,3$).  $\hat S$ is the singlet
superfield already present in the minimal NMSSM.  Besides, we have
added two more gauge singlets $\hat N$ and $\hat X$, for each
generation, which carry lepton numbers $L=-1$ and $L=+1$,
respectively. In our formulation, even though $L$ is not a good
quantum number because of the presence of a non-vanishing $\mu_X$,
$(-1)^L$ is still a good symmetry.  We have written the $\hat N \hat
X$ and $\hat X \hat X$ terms in a generation diagonal basis without
any loss of generality. Once the scalar component of
$\hat S$ acquires a VEV ($v_S$), not only the conventional $\mu$-term
is generated with $\mu = \lambda_H v_S$, a lepton number {\em
  conserving} mass term $M_{N}\Psi_{N}\Psi_{X}$ is generated as well,
with $M_{N} = \lambda_{N} v_S$. One more lepton number {\em
  conserving} mass term $m_D \Psi_{\nu} \Psi_{N}$ emerges with $m_D =
f^\nu v_u$.

The crucial term relevant for inverse seesaw is the $\Delta L = 2$
term involving $\mu_X$, which is the only mass dimensional term in the
superpotential. We assume that the $Z_3$ symmetry of the
superpotential is absent only in this term. 
We treat $\mu_X$ as an extremely tiny {\em effective} mass parameter
generated by some unknown dynamics.  Its smallness
would eventually decide the lightness of the light neutrino. We make a
few observations at this stage:

($i$)~ The superpotential treats the two singlets $\hat N$ and $\hat
X$ differently in the sense that it yields a $\mu_X \Psi_X \Psi_X$ Majorana
mass term ($\Delta L = 2$) but does not lead to a similar $\mu_N \Psi_N
\Psi_N$ term. This discrimination requires further qualification.  A
generic superpotential with $(-1)^L$ parity should have included the
latter term.  Both $\mu_N$ and $\mu_X$ can be
naturally small, as their absence enhances the symmetry of the
Lagrangian. But the important thing to note is that the magnitude of
$\mu_X$ (and not that of $\mu_N$) controls the size of the light
neutrino mass \cite{magut,dyna_inverse}. In view of this, for the sake
of simplicity, we have assumed $\mu_N = 0$.

($ii$)~ Two questions naturally arise here.  First, although we have
put $\mu_X$ by hand and claimed that it is tiny, is it possible to
dynamically generate a small $\mu_X$ starting from a superpotential
which does not {\em a priori} contain any mass dimensional term?
Second, is it possible to realize $\mu_N \ll \mu_X$ in a sensible
model?

To address the first question, we admit that with the particle content
of our model it is not possible to provide a natural solution for a
small $\mu_X$. One possibility could have been to start with a
trilinear $\lambda \hat S \hat X \hat X$ term in the superpotential of
Eq.~(\ref{model-exp}), which would lead to $\mu_X = \lambda
v_S$. Since $v_S \sim v$, the requirement to produce the correct light
neutrino mass would then compel us to take $\lambda \sim
10^{-11}$. Such a small dimensionless trilinear coupling would be
against the spirit of inverse seesaw mechanism as illustrated
earlier.  However, by extending the particle content of the model it
is possible to dynamically generate a small $\mu_X$.  In this context,
we recall that in the original inverse seesaw formulation
\cite{inverse}, the smallness of $\mu_X$ was attributed to the
supersymmetry breaking effects in a superstring inspired $E_6$
scenario. It is also possible to keep $\mu_X$ small by relating it to
a tiny VEV generated dynamically.  An analysis along this line was
carried out in an extended version of the NMSSM \cite{munoz_inverse},
where the origin of small VEV can be traced to the assumption of a
vanishing trilinear scalar coupling at the GUT scale. A small $\mu_X$
can also be realized in a supersymmetric SO(10) context \cite{devgut}.
Regarding the second question, we recall the example of a
non-supersymmetric SO(10) framework, which also contains the remnants
of a larger ${\rm E_6}$ symmetry \cite{magut}, where $\mu_X$ is
generated at two-loop level, but $\mu_N$ is generated at a higher loop
justifying its relative smallness.  In our work, we do not advocate
any specific GUT scenario to provide the dynamics that generates a
small $\mu_X$. We treat $\mu_X$ as an effective phenomenological
parameter of unspecified origin, whose smallness derives its origin in
some unknown hidden sector dynamics. We simply set its value to
reproduce the correct light neutrino mass.

We now illustrate the pattern of neutrino masses with only one
generation. In the $\{\Psi_\nu,\Psi_N,\Psi_X\}$ basis, the $(3 \times
3)$ neutrino mass matrix is given by
\begin{eqnarray}
{\cal M}&=&\left(
\begin{array}{ccc}
0 & m_D & 0 \\
m_D & 0 & M_N \\
0 & M_N & \mu_X \\
\end{array}\right) \ ,
\label{nmssm-matrix}
\end{eqnarray}
yielding  the mass eigenvalues  ($m_1 \ll m_{2,3}$):
\begin{eqnarray}
 m_1 = \frac{m_{D}^2 \mu_X}{m_{D}^2+M_{N}^2} \, , ~~~~ 
 m_{2,3} =  \mp \sqrt{M_{N}^2+m_{D}^2} + 
\frac{M_{N}^2 \mu_X}{2 (m_{D}^2+M_{N}^2)}  \, . 
\label{masses}
\end{eqnarray}
The important thing to observe here is that the lightness of the
smallest eigenvalue $m_1$ is due to the smallness of $\mu_X$. The
other two eigenvalues ($m_2$ or $m_3$) can have a mass around 10 GeV,
and their presence significantly influences the decay pattern of
$A_1$.

We now compute the branching ratios of $A_1$ into the invisible modes
comprising of the $\Psi_\nu$, $\Psi_N$ and the $\Psi_X$ states.  Rigorously
speaking, one should first diagonalize the mass matrix of
Eq.~(\ref{nmssm-matrix}) to determine the physical neutrino states. However,
for our purpose it suffices to estimate the branching fractions of $A_1$ into
the $\Psi_\nu \Psi_N$ and $\Psi_N \Psi_X$ interaction states.
Recall from Eqs.~(\ref{A1-def}) and (\ref{model-exp}) 
that the decay of $A_1$ into $\Psi_\nu \Psi_N$
will depend on how large the doublet component of $A_1$ is, i.e. on how large
$\cta$ is, whereas the decay into $\Psi_N \Psi_X$ will
depend on the amount of $A_S$ inside $A_1$, i.e. on the magnitude of $\sta$.
Below, we present the branching ratios into invisible modes normalised to the
visible ones (neglecting, for simplicity, the phase space effects).
\begin{eqnarray}
 \frac{{\Br}\left(A_1 \rightarrow \Psi_\nu \Psi_N \right)}
{{\Br}\left( A_1 \rightarrow 
f \bar f \right)+{\Br}\left( A_1 \rightarrow c \bar{c} \right)} &\simeq& 
\frac{m_{D}^2}{m_{f}^2 \tan^4 \beta + m_{c}^2} \, , \label{a1nuN}\\
 \frac{{\Br}\left( A_1 \rightarrow \Psi_N \Psi_X \right)}
{{\Br}\left( A_1 \rightarrow 
f \bar f \right)+{\Br}\left( A_1 \rightarrow c \bar{c} \right)} &\simeq& 
\tan^2\theta_A \frac{M_{N}^2}{m_{f}^2 \tan^2 \beta + m_{c}^2 \cot^{2} \beta} 
\frac{v^2}{v_S^2} \, , \label{a1sN}
\end{eqnarray}
where $v = \sqrt{v_u^2 + v_d^2} \simeq 174$ GeV. Notice that the
dominant visible decay modes of $A_1$ are $f\bar f (f=\mu, \tau, b)$
and $c \bar c$.  Of course, the $c \bar c$ mode would be numerically
relevant if $m_{A_1} < 2m_b$ and $\tanb$ is small. Note that the
branching ratio into $\Psi_N \Psi_X$ dominates over that into
$\Psi_\nu \Psi_N$ for two reasons - firstly, there is a
$\tan^2\theta_A$ prefactor for the former which can be rather large if
$A_1$ has a dominant singlet component; secondly, if the $m_f^2$ term
in the denominator of the branching ratio expressions is numerically
relevant, then the $\Psi_\nu \Psi_N$ channel suffers a suppression by
an additional $\tan^2\beta$ factor.

For a numerical illustration, we make two choices of $\tanb = (3,
20)$, and fix $\cta = 0.1$, which yield $X_d = \cta \tanb = (0.3, 2)$.
We recall that the upper limit on $X_d$ for $\ma<8~$ GeV in the
minimal NMSSM has been obtained primarily from radiative
$\Upsilon$-decays, and the limit is between 0.7 to 3.0 for $\tanb =
50$, while it is 30 or above for $\tanb = 1.5$ \cite{Domingo:2008rr}.
A value of $X_d = 2$ is in fact slightly above the upper limit for
$\ma$ in the range of 4 to 8 GeV.  In the present scenario, $A_1$ has
a significant branching ratio into invisible modes which, in turn,
considerably relax the upper bound on $X_d$. Here we do not choose a
very large value of $\tanb$ as that would increase the branching ratio
of $A_1$ into visible modes.  The value of $m_{A_1}$ is chosen to be
somewhat larger than $M_N$, so that the phase space suppression, given
by the factor $\left(\left\{1-(\frac{2m_f}{\ma})^2\right\} \Big{/}
  \left\{1-(\frac{2M_N}{\ma})^2\right\}\right)^{1/2}$, is not
numerically significant. We consider two values for $M_N = (5, 30)$
GeV. The rationale behind choosing $M_N = 5$ GeV is that it allows us
to explore $\ma <10~$ GeV, a regime where constraints from $\Upsilon$-
and $B$-decays are particularly restrictive -- see Table
\ref{cons}. On the other hand, the choice $M_N = 30$ GeV implies that
${A_1}$ is moderately heavy ($\ma >30~$ GeV) which corresponds to the
range where LEP and $B$-decay constraints are relevant.  We display
our results in Table \ref{table2}. For numerical illustration, we have
assumed $v_S \sim \mathcal O(v)$.  The main conclusion is that if
$\cta$ is small, $A_1$ has a dominant singlet component (which is
generally the case when $A_1$ is light \cite{ulrichrev}), then for a
reasonable part of the parameter space $A_1$ can have a sizable
invisible branching ratio which would weaken many of the constraints
discussed in the beginning. However, it is important to stress that
$\cta$ should not be excessively small, since in that case the purely
singlet $A_1$ would be completely decoupled from the visible sector.

\begin{table}[!ht]
\begin{center}\
\begin{tabular}{|c|c|c|c|c|c|}
\hline
\hline
&\multicolumn{2}{c|}{$\tan \beta = 20$, $\cos \theta_A = 0.1$} 
& \multicolumn{2}{c|}{$\tan \beta = 3$, $\cos \theta_A = 0.1$}\\
\hline
$M_N$ (GeV) & $5$ &$30$& $5$ &$30$ \\
\hline
\hline
${\Br}\left(A_1\rightarrow \Psi_\nu \Psi_N\right)$ 
& $ 7\times 10^{-5}$   &  $3\times 10^{-6}$   & $4\times 10^{-3}$  
&$1\times 10^{-4}$  \\
\hline 
${\Br}\left(A_1\rightarrow \Psi_N \Psi_X\right)$ 
& 0.7 & 0.9 & $\sim$ 1 & $\sim$ 1  \\
\hline
\hline
\end{tabular}
\end{center}
\caption{\em \small Invisible branching ratios of the lightest NMSSM 
  pseudoscalar for $m_D=10$ GeV, $M_N = (5, 30)$ GeV, and $\mu_X = 1$ eV.}
\label{table2}
\end{table}

What about the CP-even Higgs mass limit?  Since the $ZZh$ coupling is diluted
with respect to its SM value as a result of singlet admixture, the direct
search (lower) limit on $m_h$ from its non-observation will be lower than the
SM limit of 114 GeV. A study based on the OPAL data from the LEP-2 run shows
how the lower limit on $m_h$ decreases (assuming the Higgs production via
Higgs-strahlung process) as $\xi \equiv \sigma(Zh) {\rm Br}(h \to ~{\rm
  invisible})/\sigma^{\rm SM} (Zh)$ gets smaller than unity
\cite{Nagai:2008zz}. In our case, we have not only a mixing between the MSSM
part of the CP-even Higgs and the singlet CP-even component, but also a
sizable branching ratio of $h \to A_1 A_1 \to ~{\rm invisible}$.  The lower
limit on the lightest CP-even Higgs mass will then decrease accordingly.

To conclude, we explore the possibility of having a very light (of order 10
GeV) pseudoscalar in the NMSSM, which has interesting consequences for CP-even
neutral Higgs search at colliders, as well as for facilitating dark matter
annihilation.  In this context, we have extended the minimal NMSSM with two
additional gauge singlets carrying opposite lepton numbers for two specific
reasons. On one hand, they provide a substantial invisible decay channel to
the lightest pseudoscalar which helps relaxing or even evading some of the
tight constraints from $\Upsilon$- and $B$-decays. On the other hand, they
naturally set up the stage for implementing the inverse seesaw mechanism in
order to generate light neutrino masses. What is phenomenologically
interesting is that this can be done using order one neutrino Yukawa couplings
and employing neutrino Dirac masses of a few tens of GeV. To account for the
experimental values of the two mass squared differences and the three mixing
angles of light neutrinos, one would of course have to extend the number of
$\hat N$ and $\hat X$ superfields.

\noindent {\bf Acknowledgments:}~
We thank U.~Ellwanger and A.M.~Teixeira for some valuable comments and
suggestions. G.B. acknowledges hospitality at the LPT, Orsay (Universit\'e de
Paris-sud 11), when this work started.  D.D. acknowledges support from the
Groupement d'Int\'er\^et Scientifique P2I.  This work has been done partly
under the ANR project CPV-LFV-LHC NT09-508531.

\end{document}